\documentclass [amssymb, twocolumn, amsmath, showpacs]{revtex4-1}
\usepackage{graphicx,epsfig,amsfonts,amssymb}
\usepackage{bm}
\usepackage{color}
\usepackage{times}
\newcommand{\be}{\begin{eqnarray}}
\newcommand{\ee}{\end{eqnarray}}

\newcommand{\la}{\langle}
\newcommand{\ra}{\rangle}

\begin{document}

\title{Quantum Zeno effect as a topological phase transition in full counting statistics and spin noise spectroscopy}

\author {Fuxiang Li$^{a,b}$, Jie Ren$^{b}$, N.~A. {Sinitsyn}$^{b}$. }


\address{$^a$ Department of Physics, Texas A\&M University, College Station, TX 77840, USA}
\address{$^b$ Theoretical Division, Los Alamos National Laboratory, Los Alamos, NM 87545,  USA}

\begin{abstract}
When the interaction of a quantum system with a detector is changing from  weak to strong coupling limits, the system experiences a transition from the regime with  quantum mechanical coherent oscillations to the regime with a frozen dynamics. In addition to this quantum Zeno transition,  we show that the full counting statistics of detector signal events also experiences a topological phase transition at the boundary between two phases at intermediate coupling of a quantum system to the detector. We demonstrate that this transition belongs to the class of topological phase transitions that can be classified by elements of the braid group.  We predict that this transition can be explored experimentally by means of the optical spin noise spectroscopy.
\end{abstract}

\date{\today}
\maketitle


{\it Introduction.} Frequently repeated  measurements applied to a quantum system may suppress quantum coherent oscillations by forcing this system to
remain in an eigenstate of the measurement operator, which is the essence of the quantum Zeno effect \cite{Zeno77}. This effect has been  successfully observed experimentally, e.g. in the escape rate of trapped ultracold sodium atoms from the trapping potential \cite{Fischer01}  or in the decay of an excitation of a Bose-Einstein condensate embedded in an optical lattice \cite{Bar09}.

The weak measurement framework allows one to treat the coupling of a system to a detector as a continuous parameter, which can be varied between the limits of the week and strong coupling \cite{Peres, Wiseman, Korotkov, Katz}. One can expect that in the low coupling limit the system is only weakly perturbed and generally exhibits coherent quantum effects such as oscillations of the measurable characteristics with time \cite{Katz}. Respectively, in the limit of a strong coupling,  coherent quantum oscillations are suppressed and transitions between states of the system become possible only due to the additional coupling of the system to its environment, which generally leads to incoherent stochastic behaviors.

It has been argued previously that such a transition between the two regimes is usually marked by a critical boundary that separates phases with and without coherent oscillations \cite{Zeno-pt}. For example, suppression of coherent spin precessions due to a continuous measurement has been recently observed in a solid state qubit in the diamond \cite{diamond-Zeno}. 
In this article we explore such a phase transition from the point of view of the full counting statistics of detector signal events \cite{FCS}. We assume that in a weak measurement process, the detector output is a series of discrete ``clicks" separated by random intervals of time. Clicks correspond to successful measurements of the system  \cite{sham}. Statistics of such events can be described by a generating function, which is  akin to the one that describes statistics of observed photons in quantum optics \cite{Scully}. The main finding of this letter  is that, when the system's interaction with a detector changes between weak and strong coupling limits, the full counting statistics undergoes a topological phase transition of the type that has been introduced recently to classify band structures of non-Hermitian Hamiltonians \cite{BPT}. 

We will prove that { the cumulant generating function} and the time correlator of the detector output signal show damped oscillations in one phase and a monotonous decay with time in the other phase.
The frequency of damped oscillations is finite in one phase but approaches zero value near the critical point. At higher than critical couplings, oscillations disappear, which
makes the oscillation frequency a natural order parameter whose value distinguishes  the quantum coherent phase from the quantum Zeno phase. The observation that different phases are topologically distinct is important, in particular, to conclude that the characteristics of different phases are topologically protected, i.e. they should be conserved upon finite changes of parameters and adding small changes to the measurement model. 

{\it Setup.}
We consider a weak measurement model that was discussed in detail in \cite{sham}. It consists of a single electron spin, such as in a quantum dot or a spin-1/2 atom, which is continuously measured by means of the optical spin noise spectroscopy \cite{Oestreich05, Muller08, Crooker09, Oestreich10, Crooker10, Li12}.
The bare Hamiltonian (without dissipation sources) describes the spin precession in an in-plane magnetic field $B_y\hat{y}$:
\begin{equation}
\hat{H} = \frac{1}{2} g B_y \hat{\sigma}_y.
\label{ham}
\end{equation}
And the density matrix of spin-1/2 can be written in the form
\begin{equation}
\hat{\rho} = \frac{1}{2}  {\bm \rho} \cdot \hat{\bm \sigma}= \frac{1}{2} \left( \rho_0 \hat{1} + \rho_z  \hat\sigma_z + \rho_x  \hat\sigma_x \right)
\label{dm}
\end{equation}
with $\rho_0=1$. 
The spin rotates arount $y$-axis with Larmor frequency $\omega_L=g B_y$ so that we can have $\rho_y(t)=0$, as only incoherent relaxation is happening along $y$-axis.
We assume that spin relaxations along all axes happen with the same relaxation time $T$. As such,
dynamics of the spin with time $t$ can be described by the evolution operator $\hat{U}_t[\hat{\rho}]$ through $\hat{\rho}(t)=\hat{U}_t[\hat{\rho}(0)]$, reading explicitly:
\be
\hat{\rho}(t)&=& \frac{1}{2}  \Big{(} \rho_0\hat{1} +e^{-rt} \{ [\rho_x(0)\cos\omega_Lt+ \rho_z(0)\sin \omega_L t]\hat{\sigma}_x \nonumber\\
 && \quad + [\rho_z(0)\cos\omega_Lt-\rho_x(0)\sin\omega_Lt]\hat{\sigma}_z  \}\Big{)},
\label{eq:U}
\ee
where $r=1/T$ denotes  the relaxation rate. We will assume that $r\ll \omega_L$ so that without coupling to the detector, the spin performed weakly damped coherent precession under the magnetic field.

We further assume that measurements are performed in small discrete steps $\tau \ll T,1/\omega_L$. Hence, it is possible to develop a continuous limit of our weak measurement scheme.
The term ``weak'' means here that there is only a small probability per each measurement for the detector to collapse the state vector of the spin to a measurement operator eigenstate. Let our measurement axis be $z$-axis.
The spin detection occurs due to the Faraday rotation experienced by the linearly polarized beam \cite{Greilich, Mikkelsen, Berezovsky, Berezovsky2}. Following \cite{sham}, we will assume that the detector is tuned to
be insensitive to the beam passing through the spin state $|+\rangle$, but if the spin is in
the state $|-\rangle$, the beam experiences additional rotation of polarization, which has finite but small probability  per measuring interval $\tau$ to produce an elementary distinguishable ``click" of the detector.

Consider now the spin in a superposition
\begin{equation}
|\psi \rangle = a_+ |+\rangle +a_-|-\rangle,
\label{super}
\end{equation}
with coefficients $a_+$ and $a_-$.
After one weak measurement, with a probability $p_D|a_{-}|^2$, the detector shows a ``click", i.e. its output signal (e.g. voltage) intensity $I(t)$ shows a pulse denoted as 1, and the density matrix of the spin collapses to the pure state $|-\rangle \langle - |$. Here, the  probability $p_D \ll 1$ characterizes the capability of the detector to  induce  the collapse of the wave function per one measurement. Respectively, with the probability $(1-p_D)$, the detector will not respond, i.e. its output voltage is zero at such time intervals.
Thus, the output of such an ideal detector produces a sequence, e.g.:
\begin{equation}
I(t) \sim \ldots 00\,{\rm 1}\, 00000000\,{\rm 1}\, 000000 \,{\rm 1}\,000\, {\rm 1}\,000\ldots
\label{sequence1}
\end{equation}
The information content of a detector signal, such as in (\ref{sequence1}), can be determined from the signal statistical characteristics, which are in the focus of this article.

{\it Counting statistics of detector events.}
 Following \cite{sham}, we use POVM formalism \cite{Peres} to obtain the probability of observing the sequence, such as (\ref{sequence1}), by introducing the Kraus operators:


(i) Result  ``1"  and  projection of the density matrix on the $|- \rangle$ state are described by the Kraus operator
\begin{equation}
\hat{M}_1 = \sqrt{p_D} |-\rangle \langle - |,
\label{m2}
\end{equation}


(ii) Result ``0" does not correspond to a collapse of a state vector. It is described by the Kraus operator
\begin{equation}
\hat{M}_0 = \sqrt{1-p_D}  |-\rangle \langle - |+|+\rangle \langle + | .
\label{m2}
\end{equation}

It will be convenient to introduce parameter $\lambda_D$ such that $p_D = 4\lambda_D \tau$. In the limit $\tau \rightarrow 0$, parameter $\lambda_D$ does not depend on $\tau$. Moreover, since $p_D$ in this limit is proportional to the beam intensity, so does the parameter $\lambda_D$. Another physical meaning of this parameter (as we will show later) is the mean value of the inverse time between successive detector clicks separated by zeros.
This means that the value of $\lambda_D$ is a natural parameter that describes the strength of the coupling to the detector. At large values of $\lambda_D$ system should show a pronounced Zeno effect  and at small $\lambda_D$ coherent oscillations should be observed in characteristics of the detector output. We will explore what happens at intermediate values of $\lambda_D$ when this parameter is continuously varied.

The probability of a sequence $X\equiv [x_1\,x_2 \ldots x_n] $ as a string of binary numbers is given by
$P_X={\rm Tr} \left (\hat{M}_X [\hat{\rho}] \right)$, where
\begin{equation}
\hat{M}_X = \hat{M}_{x_n} \hat{U}_{\tau} \hat{M}_{x_{n-1}} \ldots \hat{M}_{x_2} \hat{U}_{\tau} \hat{M}_{x_1}.
\label{sequence2}
\end{equation}
with $\hat{M}_{x_n} [\hat{\rho}] \equiv \hat{M}_{x_n} \hat{\rho} \hat{M}_{x_n}$ and  $\hat{U}_t[\hat{\rho}]$ defined above Eq.~\ref{eq:U}.
Let $P_i(n,t)$ be the  $i$th component of the spin density matrix at time $t$ given that $n$ clicks have been detected during the measurement time $t$.
Such a density matrix is obtained by summing over all $\hat{M}_X (t)$ in which operator $\hat{M}_{1} [\hat{\rho}]$ encounters exactly $n$ times.
The component $\rho_y$ remains zero during the evolution while,
using Eq.~(\ref{sequence2}) and considering the continuous limit $\tau \rightarrow 0$, we obtain the equation of motion for the vector ${\bm P}(n, t) \equiv \{P_0(n, t),P_z(n, t),P_x(n, t)\}$:
\be
\frac{\partial}{\partial t}{\bm P}(n, t)=(\hat{K}_0-\hat{V}){\bm P}(n, t)+\hat{V}{\bm P}(n-1, t)
\label{rr}
\ee
with
\begin{equation}
\hat{K}_0=
\left(\begin{array}{ccc}
0 & 0 & 0 \\
0 & -r & -\omega_L \\
0 & \omega_L & -(2\lambda_D +r)
\end{array}
\right)
\label{eq:K0}
\end{equation}
and
\be
\hat{V}=\left(
    \begin{array}{ccc}
      2\lambda_D & -2\lambda_D & 0 \\
      -2\lambda_D & 2\lambda_D & 0 \\
      0 & 0 & 0 \\
    \end{array}
  \right)
\ee

The full accessible information about the system and the measurement sequence is contained in the vector of generating functions with components:
\begin{equation}
{\bm Z}(\chi, t)=\sum^{\infty}_{n=0} {\bm P}(n, t)e^{in\chi}, 
\label{genz}
\end{equation}
where $\chi$ is the counting field, conjugated to the number $n$. When $\chi=0$, $\bm Z$ reduce to the vector $\bm \rho$: $\bm Z(\chi=0, t)=\sum_n \bm P(n,t)=\bm \rho(t)=\{\rho_0(t), \rho_z(t), \rho_x(t)\}$. The evolution equation for ${\bm Z(\chi, t)}\equiv\{Z_0(\chi, t), Z_z(\chi, t), Z_x(\chi, t)\}$ can be obtained  by multiplying (\ref{rr}) by $e^{i\chi n}$ and summing over $n$:
\be
\frac{\partial}{\partial t}{\bm Z}(\chi, t)=\hat{K}(\chi){\bm Z}(\chi, t),
\label{eq:Z}
\ee
with
\be
&&~~\hat{K}=\hat{K}_0+(e^{i\chi}-1)\hat{V} \label{eq:K}\\
&&=\left(\!\! \begin{array}{ccc}
2(e^{i\chi}-1)\lambda_D  &  -2(e^{i\chi}-1)\lambda_D \!\! &\!\!  0  \\
-2(e^{i\chi}-1)\lambda_D &  2(e^{i\chi}-1)\lambda_D-r \!\! &\!\!  -\omega_L \\
0  &   \omega_L \!\! &\!\!  -(2\lambda_D +r)
\end{array}\!\!
\right)\!\!. \nonumber
\ee
We will concentrate on the generating function  $Z_0(\chi,t)=\sum_n  P_0(n,t) e^{i\chi n}$ for probabilities $ P_0(n,t)$ to observe $n$ detector clicks during time $t$.  Usually the behavior of $Z_0(\chi,t)$ is of particular theoretical interest at large total measurement time $t$.

For Markovian stochastic processes $Z_0(\chi,t)$ generally has a universal form in this limit, as:
\begin{equation}
Z_0(\chi,t) \sim e^{f(\chi)t},
\label{univ}
\end{equation}
with function $f(\chi)$ having the meaning of the cumulant generating function for the number of detector clicks. This universality follows from the fact that at large $t$ the evolution of the generating function is dominated by the largest eigenvalue of some effective Hamiltonian. 
Two of us showed in \cite{BPT} that counting statistics of classical Markovian systems can undergo topological phase transitions, with distinct phases classified by the elements of the braid group. As a result,  function $f(\chi)$ becomes non-analytic at some values of $\chi$ in topologically nontrivial phases. The generating function at such points shows oscillating behavior even in the large time limit.  Below we show that analysis of the counting statistics of the quantum mechanical system with evolution equation (\ref{eq:Z}) can be performed along essentially the same steps, revealing a braid phase transition with new physical characteristics.

{\it Quantum Zeno Effect and Braid Phase Transition.}
First we observe the analogy of Eq.~(\ref{eq:Z}) and the Schr\"odingier eqation with a non-Hermitian Hamiltonian $\hat{K}(\chi)$. Note that by definition 
$\hat{K}(\chi)=\hat{K}(\chi+2\pi)$.  One can think of eigenvalues of $\hat{K}(\chi)$ as a band structure with parameter $\chi$ playing the role of the Bloch vector.
 For a non-Hermitian Hamiltonian, the eigenvalues are generally complex. The consequence is that the periodicity of the Hamiltonian implies only the periodicity of the whole nondegenerate complex eigenspectrum but not the periodicity of each energy band as a function of $\chi$.

We explored the band structure of the operator $\hat{K}(\chi)$ numerically at different values of the parameter $\lambda_D$.
We observed that, by increasing $\lambda_D$ from small to large values, the band structure of $\hat{K}(\chi)$ passes through states with eigenvalue crossings at some values of the parameter $\chi$.
The most interesting crossing point appears at $\chi=0$ because the vicinity of the point $\chi=0$ describes the most accessible lowest cumulants of the
statistics of detector clicks.  Such a crossing point appears in our system at  $\lambda_D=\omega_L$. 

\begin{figure}
{\includegraphics[width=0.9\columnwidth]{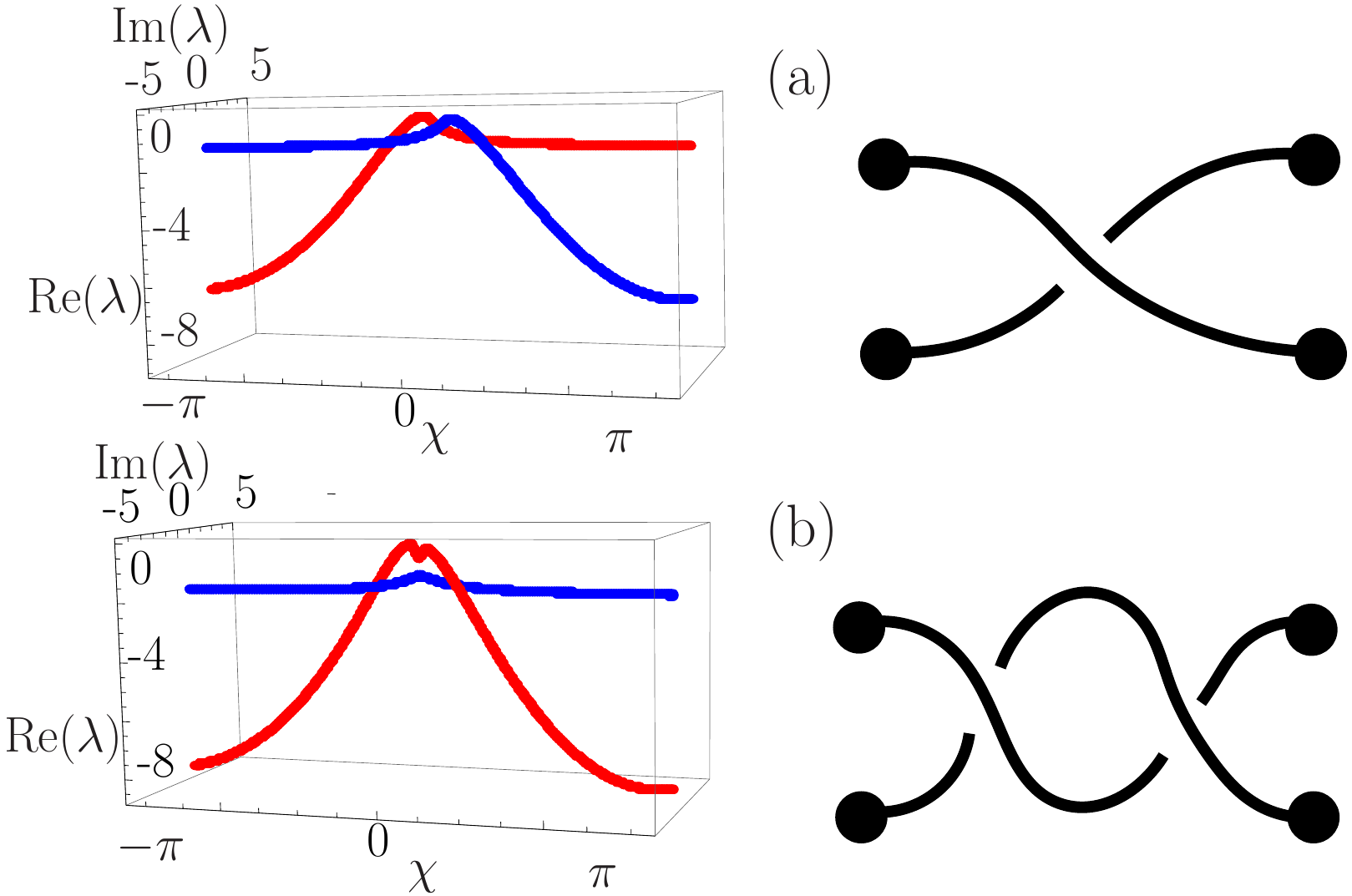}}
\hspace{-2mm}\vspace{-4mm}
\caption{{ Left:} Two  eigenvalues of operator $\hat{K}(\chi)$ having the largest real parts. Parameters are chosen as: $r=0$ and $\omega_L=1$ for (a) the phase with quantum coherent oscillations at $\lambda_D=0.7$,  and (b) the phase without coherent oscillations at $\lambda_D=1.02$. { Right}: The group elements  corresponding to topologically inequivalent band structures on the Left.}
\label{braid1}
\end{figure}

In Fig.~\ref{braid1}  we illustrate the band structure of $\hat{K}$ for $\lambda<\omega_L$ and $\lambda>\omega_L$ and the corresponding schematic representations of twisting patterns of eigenvalues in terms of the braid diagrams (for clarity, we omit one irrelevant band). For the case (a), it can be seen that each single band doesn't maintain the periodicity as does $\hat{K}(\chi)$. Then
the two displayed eigenvalues of $\hat{K}(\chi)$ satisfy the conditions:
  \begin{eqnarray}
\lambda_1(\chi)&=&\lambda_2(\chi+2\pi), ~~~~ \lambda_2(\chi)=\lambda_{1}(\chi+2\pi).
  \end{eqnarray}



 In the case (a), the two complex bands twist with each other forming an element of the braid group. When we adjust the system parameters to make 
 $\lambda_D>\omega_L$, the periodicity is restored for each single band, and the original two complex bands twist twice with each other forming a 
 different element of the braid group. The band structures for the two cases, as shown in Fig. \ref{braid1}, are 
 topologically inequivalent. Therefore, we encounter a topological phase transition. 

Braid transitions correspond to emergence or disappearance of certain oscillating modes in the full counting statistics \cite{BPT}.
Indeed,
 let 
 $\epsilon_j$ be the eigenvalues of $\hat{K}(\chi)$.  The evolution of the $Z_0(\chi, t)$ would be $Z_0(\chi, t)=\sum_{j}A_{j}(\chi)e^{\epsilon_j (\chi) t}$ with some coefficients
$A_{j}$ that depend on the initial state of the system. 
At large time $t$, behavior of the generating function at a given value of $\chi$ is dominated by eigenvalue(s) of $\hat{K}(\chi)$ having the largest real part (besides the omitted zero mode $\epsilon_0$ in the present model).
 When $\lambda_D <\omega_L$, there are two such complex conjugate eigenvalues that determine behaviors of the generating function near $\chi=0$ (explicitly, $\epsilon_{1,2}=-r-\lambda_D\pm i \sqrt{\omega_L^2-\lambda_D^2}$ at this point). 
  On the other hand, when the coupling to the detector is stronger than it is at the phase transition value, i.e. $\lambda_D>\omega_L$, only a single real eigenvalue of the operator $\hat{K}$ dominates behavior of generating function near $\chi=0$ at large time, which corresponds to the  monotonous decay with time.

To obtain a better intuition about physical consequences, consider the cumulants of the distribution of the number  $n$ of detected clicks:
\be
c_1 \equiv \langle n \rangle = \frac{\partial Z_0(\chi)}{\partial (i\chi)} {\Big |}_{\chi=0} ,
\label{cumulants}
\ee
\be
c_2 \equiv \langle n^2 \rangle-\langle n \rangle^2 = \frac{\partial^2 Z_0(\chi)}{\partial (i\chi)^2} {\Big |}_{\chi=0} -c_1^2.
\label{cumulants2}
\ee
One can obtain evolution equations for $\langle n \rangle$ and $\langle n^2 \rangle$ by differentiating Eq.~(\ref{eq:Z}) over $\chi$ once and twice and setting $\chi$ to zero.
Integrating them with equilibrium initial conditions $Z_0(\chi=0,t=0)=1$, $Z_x(\chi=0,t=0)= Z_z(\chi=0,t=0)=0$, $c_1(t=0)=0$ and  $c_2(t=0)=0$, we find
\be
c_1(t)=2\lambda_D t,
\label{c1t}
\ee
i.e. the average number of clicks just linearly increases. However, for $c_2(t)$ we find an exponentially decaying correction in addition to a linearly growing
contribution. A particularly simple expression for this component appears after we take 2nd derivative of $c_2(t)$ over time:
\be
\frac{\partial^2 c_2(t)}{\partial t^2}= 8\lambda_D^2 \left( e^{\hat{ Q}_0 t} \right)_{11},
\label{c2t}
\ee
where
\be
\hat{ Q}_0 =\left(
\begin{array}{cc}
-r & -\omega_L \\
 \omega_L & -(2\lambda_D +r)
\end{array}
\right)
\label{q0}
\ee
is the nonzero 2$\times$2 sub-matrix
of the matrix $\hat{K}_0$.
At $\lambda_D >\omega_L$, i.e. at  strong coupling to the detector, matrix $\hat{Q}_0$ has two real eigenvalues $\epsilon_{1,2}$, which correspond to monotonous decay of (\ref{c2t}) with only one of eigenvalues dominating the
decay at large time.
In contrast, at $\lambda_D <\omega_L$, i.e. at the weak coupling, the matrix $\hat{Q}_0$ has two complex conjugated eigenvalues which correspond to damped oscillating behavior of (\ref{c2t}).
Similar oscillating behavior is found in all higher order cumulants of the click distribution.
Oscillation frequency is given by the imaginary part of such an eigenvalue: $\omega=\sqrt{\omega_L^2-\lambda_D^2}$. It gradually decreases with increasing $\lambda_D$ and becomes zero at the phase transition point. Hence $\omega$ is the natural order parameter that distinguishes two phases.


Finally, we discuss the possibility to explore this phase transition experimentally.
Measurements of individual physical detector events can be a very hard task. Instead, we suggest to use the recently developed method of
 the optical spin noise spectroscopy, which allows one to measure the intensity correlator of the detector output signal at equilibrium:
\be
C_2(t)=\la I(t) I(0) \ra-\la I(t) \ra \la I(0) \ra,
\label{corr}
\ee
where the signal $I(t)$ is  given by a sequence of physical detector clicks, such as (\ref{sequence1}). Importantly, spin noise spectroscopy can effectively extract the physical correlator (\ref{corr}) from a signal with a considerable background noise even when it is impossible to resolve individual useful detector clicks  \cite{Oestreich05, Muller08, Crooker09, Oestreich10, Crooker10, Li12}. 
\begin{figure}
{\includegraphics[width=0.85\columnwidth]{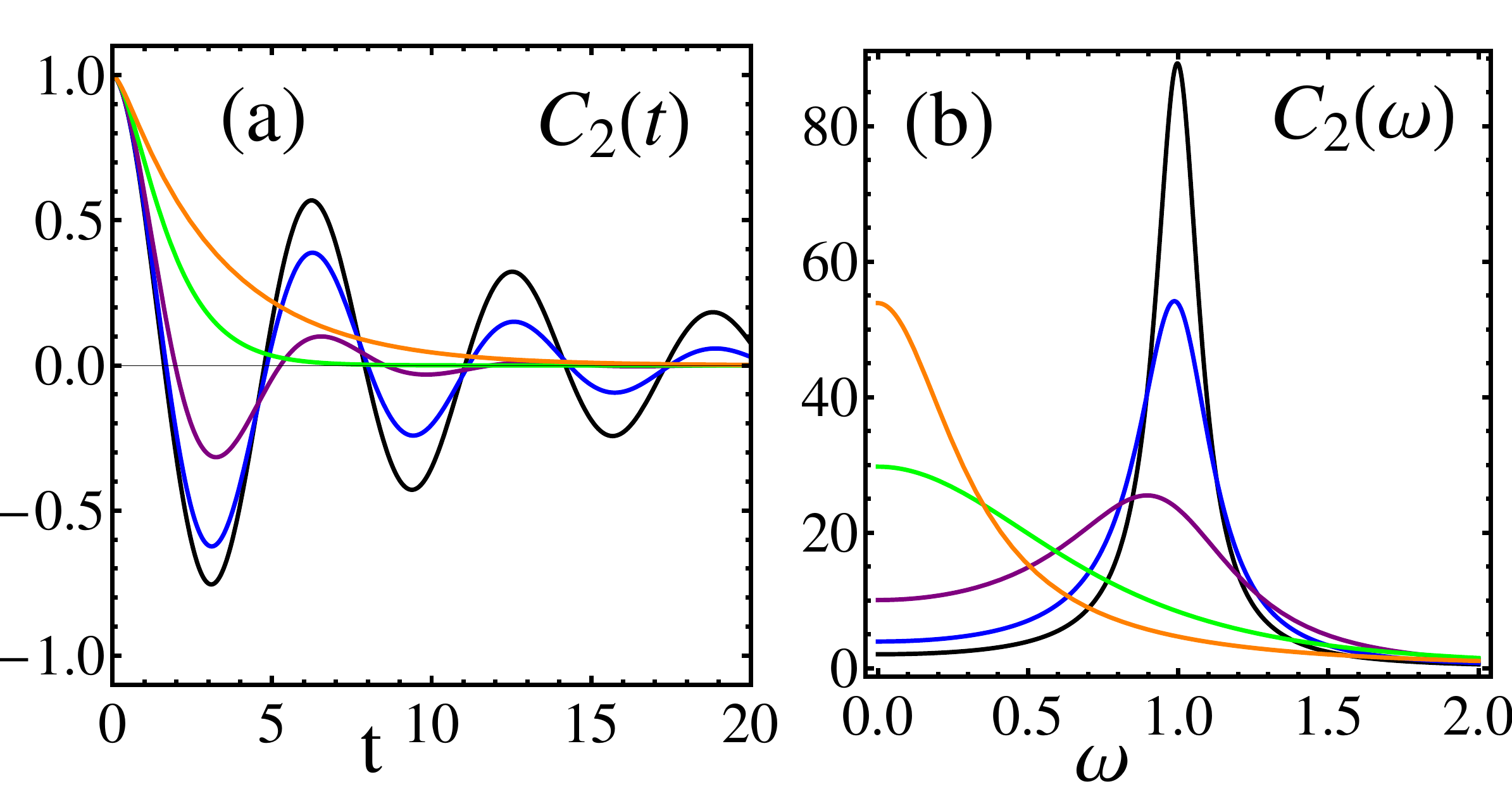}}
\hspace{-2mm}\vspace{-4mm}
\caption{(a) The correlator $C_2(t)$ as a function of time at different values of the parameter $\lambda_D$. (b) The Fourier transform of $C_2(t)$ (the noise power spectrum) at the same values of parameters in (a).   Black, blue, purple, green and orange curves correspond to, respectively,  $\lambda_D=0.04, 0.1, 0.3, 1.0, 2.0$. The Larmor frequency is $\omega_L=1$.}
\label{C2}
\end{figure}

If we know the spin density matrix at time $t$ after the system was successfully
measured to be at state $|-\ra$, then the intensity correlator can be expressed as
\be
C_2(t)=-4\lambda_D^2 \rho_z(t).
\label{c2-corr}
\ee
Evolution equation for $\hat{\rho}(t)$ can be found by noticing that ${\bm \rho(t)}={\bm Z}(\chi=0,t)$. Hence operator $\hat{K}(\chi=0,t)$ is the operator of evolution for the components $\rho_i$, defined in (\ref{dm}).
Only its 2$\times$2 sub-matrix $\hat{ Q}_0 $ is nonzero in this case so that $\rho_z(t)=\left[  -\exp (\hat{Q}_0t) \right]_{11}$, i.e. we obtain a relation:
\be
2C_2(t)=\frac{\partial^2 c_2(t)}{\partial t^2}.
\label{corr-21}
\ee
Consequently, the braid phase transitions are also directly responsible for the qualitative change of the behavior of the intensity correlator $C_2(t)$ measured at the steady conditions. At $\lambda_D< \omega_L$ correlator $C_2(t)$ shows damped oscillations that
continue for arbitrary time, while at   $\lambda_D> \omega_L$ the correlator $C_2(t)$ monotonously decays with time, as we illustrate in
 Fig.~\ref{C2}(a). In  Fig.~\ref{C2}(b), for convenience, we also show behavior of this correlator in the frequency domain.

Spin noise correlators at equilibrium are particularly convenient to study in atomic gases by measuring the spin noise power spectrum  \cite{atomic}. 
 Experimentally, parameter $\lambda_D$ can be varied by changing the intensity of the measurement beam. To achieve this, the Larmor frequency should be sufficiently small, i.e. $\omega_L \sim 1/T$, as in Fig.~\ref{C2}(b). 
We predict that varying intensity of the beam one can observe a transition from damped oscillations of the spin-spin correlator  in real time to its monotonous decay.  The major challenge is to achieve the Zeno effect regime at strong intensities of the beam that however do not substantially heat the system and vary relaxation rates. For this reason one should choose systems with  large relaxation times $T$, so that the values of $\lambda_D$ and $\omega_L$ are relatively small.  A possible candidate system is  e.g. a $^{85}$Rb atomic gas, in which the spin noise power has been studied at  mG magnetic fields values with the relaxation rate $1/T$  of only several kHz at  112$^{\circ}C$   \cite{atomic}. It is possible that by increasing the measurement beam intensity, the transition to the Zeno regime will be achieved in this system without affecting values of basic system parameters.





{\it In conclusion}, we have showed that the path between Zeno effect and  quantum coherent dynamics in the weak measurement framework is marked by a topological phase transition at an intermediate value of a system coupling to the detector.
 Different phases correspond to different topologically nontrivial braid group elements, which classify the band structure of a non-Hermitian Hamiltonian that governs the evolution of the moment generating function of the counting statistics.
Oscillations of cumulants of detector clicks at  low coupling  is the signature of quantum coherent regime, while lack of such oscillations in the phase with large coupling can be interpreted as the on-set of the quantum Zeno effect. The oscillation frequency is the order parameter distinguishing between two phases.
The above discussed phase transition could be observed in atomic gases by means of the optical spin noise spectroscopy.

{ Phase transitions at fluctuation level are critical phenomena that can be observed only in higher than the first order cumulants of the event counting statistics. They have attracted considerable attention recently \cite{rare-pt, ivanov} but their experimental studies in condensed matter systems have been complicated because most of such phenomena can be observed only on the level of extremely rare and unusual events. Our results prove, in particular, that some of such critical phenomena can be observed on the level of the easily accessible spin-spin correlator measurements. Moreover, the existing experimental results on the detection of the Zeno effect  can, in fact, be reinterpreted in terms of such phase transitions.}

\end{document}